\documentclass[aps,showpacs,graphicx]{revtex4}%,,preprint,twocolumn
\usepackage{graphicx}
\begin{document}

\title{Stable and deterministic quantum key distribution based on  differential phase shift\footnote{Published in
Int. J. Quant. Inform. 7 (2009) 739-745}}
\author{ Bao-Kui Zhao$^{1,2,3}$, Yu-Bo Sheng$^{1,2,3}$,
 Fu-Guo Deng$^{4}$\footnote{Author to whom correspondence should be addressed; Electronic mail: fgdeng@bnu.edu.cn},
 Feng-Shou Zhang$^{1,2,3}$, and Hong-Yu Zhou$^{1,2,3}$}
\address{$^1$The Key Laboratory of Beam Technology and Material
Modification of Ministry of Education, Beijing Normal University,
Beijing 100875,  China\\
$^2$Institute of Low Energy Nuclear Physics, and Department of
Material Science and Engineering, Beijing Normal University,
Beijing 100875, China\\
$^3$Beijing Radiation Center, Beijing 100875,  China\\
$^4$Department of Physics, Applied Optics Beijing Area Major
Laboratory, Beijing Normal University, Beijing 100875, China }
\date{\today }

\begin{abstract}
We present a stable and deterministic quantum key distribution
(QKD) system based on differential phase shift. With three
cascaded Mach-Zehnder interferometers with different arm-length
differences for creating key, its key creation efficiency can be
improved to be 7/8, more than other systems. Any birefringence
effects and polarization-dependent losses in the long-distance
fiber are automatically compensated with a Faraday mirror. Added
an eavesdropping check, this system is more secure than some other
phase-coding-based QKD systems. Moreover, the classical
information exchanged is reduces largely and the modulation of
phase shifts is simplified. All these features make this QKD
system more convenient than others in a practical application.
\end{abstract}
 \maketitle

\section{introduction}

Quantum key distribution (QKD)\cite{bb84} supplies a novel way for
generating a private key securely between two legitimate users, say
the sender Alice and the receiver Bob. Its security is based on the
laws in quantum mechanics such as noncloning theorem, coherence of
entangled systems, and quantum measurement, but not the computation
difficulty with a limited computation power. As an unknown quantum
state cannot be cloned \cite{nocloning}, the action done by a
vicious eavesdropper, say Eve will inevitably disturb the quantum
system and leave a trace in the outcome obtained by the two
legitimate users. Alice and Bob can detect the eavesdropping by
analyzing the error rate of a subset of instances chosen randomly.
Since Bennett and Brassard published an original protocol in 1984
(BB84)\cite{bb84}, QKD attracts a great deal of attention
\cite{rmp,longqkd,Hwang,ABC,CORE,lixhqkd08}.

The experimental implementation of long-distance QKD over an optical
fiber channel requires the two legitimate users to control the
influence of the fluctuation of the birefringence  which alters the
polarization state of photons. For overcoming this noise, several
elegant QKD schemes have been developed with some unbalanced
Mach-Zehnder interferometers (MZIs), such as the QKD scheme base on
the phase difference of single photon \cite{b92,phaseguo,dps}, the
"plug and play" system \cite{plugplay1} and its modifications
\cite{plugplay2,plugplay3,plugplay4,plugplay5}, the QKD system based
on faithful qubit distribution with additional qubits
\cite{yamamoto}, and the QKD system with faithful single-qubit
transmission \cite{lixhapl}. With the development of technology, the
QKD system based on faithful qubit distribution \cite{yamamoto}
seems to be perfect if there is an ideal single-photon source which
can produce two photons in a deterministic time, although the
success probability of this system for BB84 protocol is no more than
1/16 in principle in a passive way. The QKD system with the faithful
single-qubit transmission technique \cite{lixhapl} is more efficient
than that in Ref.\cite{yamamoto} as it only requires one photon in
each signal time and the qubit transmitted can reject the error
arisen from collective noise by itself. Its success probability for
generating a private key is 1/4 in theory with BB84 protocol in an
absolutely passive way. With some cascaded unbalanced Mach-Zehnder
interferometers (MZIs), a special encoder and a special decoder
\cite{deng}, one-way QKD can be implemented against collective noise
(with which the fluctuation is so slow in time that the alteration
of the polarization is considered to be the same over the sequence
of several photons or wavepackets \cite{yamamoto}) in a passive way
with a success probability approaching the intrinsic one in BB84 QKD
protocol \cite{bb84}.

In those QKD protocols with phase coding
\cite{b92,dps,phaseguo,plugplay1,plugplay2,plugplay3,plugplay4,plugplay5},
the two legitimate users need not share a reference frame for
choosing the common polarization bases, which makes these protocols
more convenient than those with polarization coding in a practical
application. The influence of the birefringence effect in fibers on
the one-way QKD systems \cite{b92,phaseguo,dps} is far more severe
than that on the two-way "plug and play" QKD systems
\cite{plugplay1,plugplay2}. Thus "plug and play" QKD systems
\cite{plugplay3,plugplay4,plugplay5} based on differential phase
shift (DPS) \cite{dps} were proposed for increasing the key creating
efficiency with some unbalanced MZIs and Faraday mirrors. These
systems are stable as the Faraday mirrors can be used to
automatically compensate for birefringence and
polarization-dependent losses in the transmission fiber
\cite{plugplay4}. Although these systems
\cite{plugplay3,plugplay4,plugplay5} are better than some others,
there are some spaces for improving. First, most faint laser pulses
split by the front Mach-Zehnder interferometer (MZI) are combined
again by the next ones, which makes the key creation efficiency
improved be limited. Second, in order to compensate for the
amplitude differences caused by the overall interference of pulses
travelling through different paths, phase shifters should be
inserted in the long arms of MZIs, which increases the complexity
and decreases stability of the system. Moreover, as there is no
eavesdropping check process in the first transmission, the systems
are vulnerable to an eavesdropping technique known as the
intercepting-resending attack \cite{rmp}. As pointed out in Ref.
\cite{lixhpra}, for each block of transmission, an eavesdropping
checking is inevitable for secure communication no matter what is
transmitted with a quantum channel.

In this paper, we propose a stable and deterministic QKD system
based on DPS with a high key creation efficiency approaching 100\%
by using the least cascaded MZIs with different arm-length
differences. Compared with that in Ref.\cite{plugplay4}, the key
generation efficiency is $7/8$ in principle, more than $3/4$, when
the two parties exploit three MZIs for creating their private key.
Its security is much higher than the latter. Moreover, this system
is stable and deterministic. Any birefringence effects and
polarization-dependent losses in the long-distance fiber are
automatically compensated with a Faraday mirror. The modulation of
the phase shifts is also more easier than others, and the
classical information exchanged is reduced largely.

\section{stable and deterministic QKD system}

Fig.1 shows the setup of our QKD system based on  DPS. It is made
up of three cascaded MZIs (MZI$_1$, MZI$_2$, and MZI$_3$) in Bob's
site for creating a private key, and a Faraday mirror and a phase
modulator in Alice's site. Similar to Ref.
\cite{plugplay4,plugplay5}, the MZIs, with long and short arms
connected by 50\%:50\% fiber couplers C$_1$ - C$_5$, are designed
to have different arm-length differences so that each pulse split
by the front MZI does not overlap with others in the process of
preparing the quantum signal. The time delays between the long and
the short arm of MZI$_1$, MZI$_2$, and MZI$_3$ are 4T, 2T, and T,
respectively. The Faraday mirror is used to  automatically
compensate for birefringence and polarization-dependent losses in
the transmission fiber. The phase modulator is used to encode the
key.

\begin{figure}[!h]
\begin{center}
\includegraphics[width=8cm,angle=0]{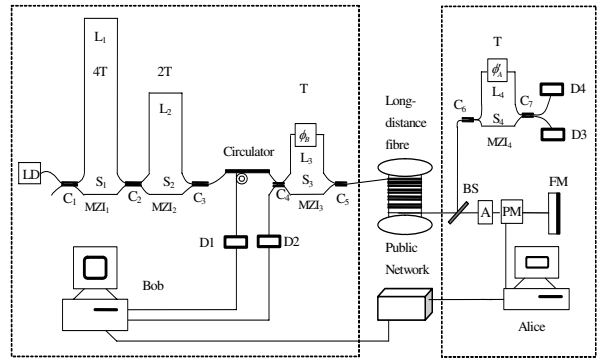}
\caption{Stable and deterministic DPS-based QKD system with a key
creation efficiency of 7/8. LD represents laser diode. C$_1$-C$_7$
represent 50\%:50\% couplers. MZI$_1$, MZI$_2$, MZI$_3$, and
MZI$_4$ are four Mach-Zehnder interferometers with different
arm-length differences 4T, 2T, T, and T, respectively. L$_1$-L$_4$
and S$_1$-S$_4$ are the long (L) arms and the short (S) arms of
MZIs, respectively. D1-D4 are four avalanche photon detectors. PM,
FM, and A represent a phase modulator, a Faraday mirror, and an
attenuator, respectively. For eavesdropping check, Alice samples a
subset of quantum signals with beam splitter (BS) and measured
them with another MZI (MZI$_4$) by choosing two bases ($\phi'_A\in
\{0, \pi/2\}$).}\label{f1}
\end{center}
\end{figure}

An original pulse  $\psi_1=e^{-i\phi_0}\vert t_1\rangle$ from a
laser diode  is split into two sequential pulses with a time
interval 4T by passing through  MZI$_1$, i.e.,
$\psi_2^1=e^{-i\phi_0}\vert t_1\rangle$ and
$\psi_2^2=e^{-i\phi_0}\vert t_5\rangle$ are combined at the second
coupler $C_2$ at time instances $t_1$ and $t_5$. Where $\phi_0$ is
the initial phase factor and the subscript 2 in $\psi_2^i$ is used
to label the second coupler. We neglect the global factor in each
pulse in this paper. Subsequently, each of the pulses is split
into two sequential pulses by MZI$_2$ with a time interval 2T,
i.e., there are four pulses described as
\begin{eqnarray}
\psi_3^1 &=& e^{-i\phi_0}\vert t_1\rangle,\;\;\;\;
\psi_3^2 = e^{-i\phi_0}\vert t_3\rangle, \nonumber\\
\psi_3^3 &=& e^{-i\phi_0}\vert t_5\rangle,\;\;\;\;
\psi_3^4=e^{-i\phi_0}\vert t_7\rangle. \label{state1}
\end{eqnarray}
After MZI$_3$ with a time interval T, there are eight sequential
pulses arriving at the coupler C$_5$. These eight pulses have the
same amplitude and the delay between two near pulses  is T (shown
in Fig.2). These eight pulses can be described as
\begin{eqnarray}
\psi_5^1 &=& e^{-i\phi_0}\vert t_1\rangle,\;\;\;\;
\psi_5^2 = e^{-i(\phi_0+\phi_B)}\vert t_2\rangle, \nonumber\\
\psi_5^3 &=& e^{-i\phi_0}\vert t_3\rangle,\;\;\;\;
\psi_5^4=e^{-i(\phi_0+\phi_B)}\vert t_4\rangle, \nonumber\\
\psi_5^5 &=& e^{-i\phi_0}\vert t_5\rangle,\;\;\;\;
\psi_5^6 = e^{-i(\phi_0+\phi_B)}\vert t_6\rangle, \nonumber\\
\psi_5^7 &=& e^{-i\phi_0}\vert t_7\rangle,\;\;\;\;
\psi_5^8=e^{-i(\phi_0+\phi_B)}\vert t_8\rangle, \label{state2}
\end{eqnarray}
where $\phi_B$ is a phase shift added by Bob for preparing the
quantum signal. Bob chooses randomly one  of the four phase shifts
$\{0, \pi/2, \pi, 3\pi/2\}$ for each original pulse
$\psi_1=e^{-i\phi_0}\vert t_1\rangle$  in the quantum
communication. That is, Bob chooses four nonorthogonal
phase-coding states to carry the message transmitted, which is
similar to the way in polarization coding that the two legitimate
users choose four nonorthogonal states $\{\vert 0\rangle, \vert
1\rangle, (\vert 0\rangle + \vert 1\rangle)/\sqrt{2},  (\vert
0\rangle - \vert 1\rangle)/\sqrt{2}\}$ to complete a deterministic
quantum communication \cite{QOTP,bidqkd}. This feature can forbid
an eavesdropper to eavesdrop the QKD system freely with
intercepting-resending attack \cite{rmp}.

\begin{figure}[!h]
\begin{center}
\includegraphics[width=8cm,angle=0]{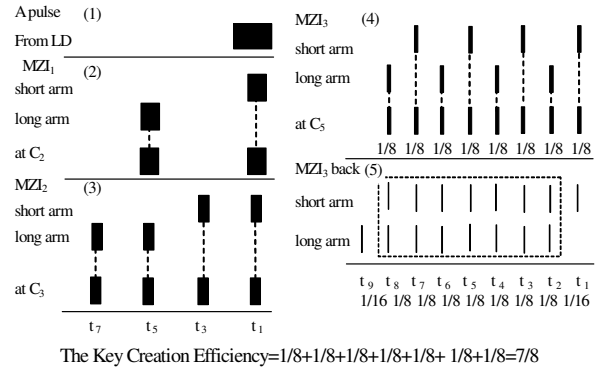}
\caption{ Process of the key creation. A pulse from the LD is
split into two pulses after travelling through MZI$_1$, and
MZI$_2$ and MZI$_3$ split them into four and eight pulses,
respectively. When these pulses are reflected back after coding,
7/8 of the pulses interfere at the coupler C$_4$ and can be used
to generate the private key, higher than $3/4$ in Ref.
\cite{plugplay4}. }\label{f2}
\end{center}
\end{figure}

When the eight pulses arrive at Alice's site, she first attenuates
the signal with a variable attenuator and then encodes her random
key on the odd pulses with the same phase shift $\phi_A \in \{0,
\pi\}$. 0 and $\pi$ represent the bit values in key string 0 and
1, respectively. After the coding performed by Alice, the quantum
signal becomes
\begin{eqnarray}
\psi_A &=& e^{-i(\phi_0 + \phi_A)}\vert t_1\rangle +
e^{-i(\phi_0+\phi_B)}\vert t_2\rangle + e^{-i(\phi_0 +
\phi_A)}\vert
t_3\rangle \nonumber\\
 &+& e^{-i(\phi_0+\phi_B)}\vert t_4\rangle +
e^{-i(\phi_0 + \phi_A)}\vert  t_5\rangle +
e^{-i(\phi_0+\phi_B)}\vert
t_6\rangle   \nonumber\\
 &+& e^{-i(\phi_0 + \phi_A)}\vert t_7\rangle +
e^{-i(\phi_0+\phi_B)}\vert t_8\rangle. \label{state3}
\end{eqnarray}
Alice reflects the quantum signal to Bob by a Faraday mirror, same
as Ref.\cite{plugplay1,plugplay2,plugplay3,plugplay4,plugplay5}.

When the pulses are reflected to Bob's site and pass through the
MZI$_3$ and the coupler $C_4$, their state becomes
\begin{eqnarray}
\psi_B &=& e^{-i\phi_0}\{e^{-i\phi_A}\vert t_1\rangle +
e^{-i\phi_B}(1 + e^{-i\phi_A})(\vert t_2\rangle + \vert t_4\rangle
\nonumber\\
&+& \vert t_6\rangle + \vert t_8\rangle) + (e^{-2i\phi_B} +
e^{-i\phi_A})(\vert
t_3\rangle + \vert t_5\rangle\nonumber\\
&+& \vert t_7\rangle) + e^{-2i\phi_B}\vert t_9\rangle\}.
\label{state4}
\end{eqnarray}
Bob can read out the key with success probability $7/8$ by using
two detectors $D_1$ and $D_2$. When $\phi_A=0$, the pulse train
clicks the detector $D_1$ at the time instances $t_2$, $t_4$,
$t_6$, or  $t_8$; Otherwise, it clicks the detector $D_2$ at one
of these four time instances. At the other three time instances
$t_3$, $t_5$, and $t_7$, the detector clicked depends on both the
phase shifts $\phi_A$ and $\phi_B$, shown in Eq.(\ref{state4}). In
detail, when $2\phi_B\oplus 2\pi =\phi_A$, the pulse train clicks
the detector $D_1$ at the time instances $t_3$, $t_5$, or $t_7$;
Otherwise, it clicks the detector $D_2$. In a word, Bob can obtain
the key with his phase shift $\phi_B$ at the time instances $t_2$,
$t_3$, $t_4$, $t_5$, $t_6$, $t_7$, or $t_8$ in a deterministic
way. At the time instances $t_1$ and $t_9$, no interference takes
place and the pulse train will click one of the two detectors
randomly, which happens with the probability $1/8$. Bob discards
these useless time instances. In quantum communication, Bob need
only tell Alice the fact that he detects a photon or not in the
useful time instances, which will reduce the classical information
exchanged largely, compared with that in
Ref.\cite{plugplay4,plugplay5} as Bob does not announce the
detailed time slots.

In order to prevent Eve from eavesdropping with
intercepting-resending attack strategy, we add a phase shift $\phi_B
\in \{0, \pi/2, \pi, 3\pi/2\}$ in Bob's site. Moreover, a
eavesdropping check is designed in Alice's site. That is, Alice
should sample some quantum signals randomly and measure them with
two nonorthogonal phase bases, shown in Fig.1, same as
Ref.\cite{QOTP,bidqkd}. Without these two tricks, this DPS QKD
protocol is insecure. We give the detail of a special
intercepting-resending attack strategy to demonstrate the necessity
of the procedure of eavesdropping check. This strategy works in the
protocols in Refs.\cite{plugplay3,plugplay4,plugplay5}, which means
these two QKD protocols is insecure in principle. For obtaining the
phase shifts performed by Alice, Eve first intercepts all the eight
pulses travelling through the coupler $C_5$ and then stores them.
She prepares another pulse $a$ and splits it into two parts, $b$ and
$c$. For evading the energy check done by Alice, Eve controls the
intensity of her pulses b and c to be equal to each of the eight
pulses $\vert t_1\rangle$-$\vert t_8\rangle$ sent by Bob. Eve sends
the part c to Bob, instead of each of the eight pulses. No matter
what the phase shift $\phi_A \in \{0,\pi\}$ is chosen by Alice, Eve
can in principle get this information by interfering the part $b$
with $c$ when it is reflected from Alice. In this way, Eve can
pretend Alice and encodes Alice's phase shift on the original eight
pulses and resends them to Bob. This attack will in principle leave
nothing in the outcome obtained by the two legitimate users. The
same way can be used to attack the quantum communication in
Refs.\cite{plugplay3,plugplay4,plugplay5}. For example, for the
pulse train $P_1$ in Ref. \cite{plugplay4}, Eve replaces the
original one with the part $c$ and sends it to Alice. No matter what
the phase shift is chosen by Alice from the two values $\{\pi/3,
4\pi/3\}$, Eve need only add a phase shift $-2\pi/3$ on the part $b$
and then interfere it with the part $c$ reflected by Alice.
Obviously, Eve can determine that the phase shift performed by Alice
is $\pi/3$ or $4\pi/3$ in principle. The phase shifts on the other
three pulse train $P_2$, $P_3$, and $P_4$ can also be obtained in
the same way. In essence, this insecurity comes from the lack of the
eavesdropping check done by Alice. As pointed out in Ref.
\cite{lixhpra}, for each block of transmission, an eavesdropping
checking is inevitable for secure communication no matter what is
transmitted with a quantum channel. This principle can be used to
ensure the security of our DPS QKD system. If the influence of the
birefringence effect in fiber is large enough, Alice should prepare
a subset of nonorthogonal pulses with which she replace some signal
pulses with a probability $p_d$ for preventing Eve from measuring
the pulses reflected by Alice. She can produce her nonorthogonal
pulses by randomly adding one of the two phase shifts $\phi_A \in
\{0, \pi/2\}$ on some signal pulses. In this time, Bob should
modulate her another phase shift $\phi'_B\in \{0, \pi/2\}$ besides
$\phi_B$ with a small probability. For improving the key creation
rate, Alice and Bob can use two biased phase bases to prepare and
measure the nonorthogonal pulses, same as that in Ref. \cite{ABC}.

\section{summary}

In summary, we have proposed a stable and deterministic QKD system
based on DPS.  The most specific feature of this DPS QKD system is
the high key creation efficiency and the high security. By using
three MZIs with different arm-length differences for creating key
(the fourth MZI is used to check eavesdropping), the efficiency is
improved to be $7/8$, more than $3/4$ in Ref. \cite{plugplay4}.
Moreover, this system is expansible. With n MZIs for creating the
key, the key creation efficiency can be added up to
$\frac{2^{n}-1}{2^{n}}$, higher than $\frac{n}{n+1}$ in Ref.
\cite{plugplay4} (much higher than that in Ref.\cite{plugplay5}).
When n is large enough, the efficiency approaches 100\%. Another
feature is the high security. With eavesdropping check and
nonorthogonal pulses, this QKD system is obviously more secure than
that in Ref. \cite{plugplay4}. Moreover, this system is stable and
deterministic, and the classical information exchanged is reduced
largely.

\section*{Acknowledgments}

This work is supported by the National Natural Science Foundation
of China under Grant No. 10604008, A Foundation for the Author of
National Excellent Doctoral Dissertation of China under Grant No.
200723, and  Beijing Natural Science Foundation under Grant No.
1082008.

\end{document}